\documentclass[final,3p,times]{elsarticle}

\usepackage{amssymb}
\usepackage{multirow}
\usepackage{amsthm}

\journal{Chinese Journal of Physics} 

\begin{document}

\begin{frontmatter}



\title{Charged analogues of singularity-free anisotropic compact stars under linear $f(Q)$-action}

\author[inst1]{Debadri Bhattacharjee}
\author[inst1]{Pradip Kumar Chattopadhyay}
\affiliation[inst1]{organization={IUCAA Centre for Astronomy Research and Development (ICARD), Department of Physics, Coochbehar Panchanan Barma University, Vivekananda Street, District: Coochbehar,  Pin: 736101, West Bengal, India.}}
\begin{abstract}
This study simulates the characteristics of spherically symmetric, anisotropic compact stellar bodies with electrical charge within the framework of the $f(Q)$ theory of gravity. Employing the Krori-Barua metric ansatz (K.D. Krori, J. Barua, J. Phys. A: Math. Gen. 8 (1975) 508) along with a linear form of $f(Q)$ model, {\it viz.}, $f(Q)=\alpha_{0}+\alpha_{1}Q$, we obtain a tractable set of exact relativistic solutions of the field equations. A specific form of charge $(q=q_{0}r^{3})$ is considered here for the present analysis. It is noted that the model is valid up to the value of charge intensity $q_{0}\leq0.0009~Km^{-2}$. Beyond this value, the model does not permit physically viable results. We have obtained the best fit equation of state in the model, which is incorporated to solve the TOV equations numerically to determine the mass-radius relation within the parameter space used here. With increasing charge intensity $(q_{0})$ from 0.0002 to 0.0009, the maximum mass ranges from $2.84-2.92~M_{\odot}$, and the corresponding radii range from $12.00-12.20~Km$. Moreover, the predicted radii of some recently observed pulsars and GW 190814 show that our model also complies with the estimated radii based on the observational results. Our model is found to satisfy all the characteristic features, such as behaviour of matter variables, causality condition, energy constraints and stability criteria, which are pertinent in the context of a stable stellar configuration to emerge as a viable and physically acceptable stellar model in the framework of $f(Q)$ gravity.
\end{abstract}




\end{frontmatter}



\section{Introduction}\label{sec1} 
Einstein's exploration of static gravitational fields led him to a daring proposition. A straightforward application of his principle of equivalence, combined with the fundamental outcomes of special relativity, implied that the gravitational field is characterised by the metric tensor. He hypothesised that this holds true even beyond the static limit and this set him on a three-year quest that concluded in November 1915 with the formulation of the field equations of his General Theory of Relativity (GR). In the later years, GR has solved many persistent mysteries pervading in astrophysics and cosmology. Despite being remarkably successful, GR has faced significant challenges when confronted with astrophysical and cosmological observational results. For instance, numerous distinct cosmological studies suggest that our universe is undergoing an accelerated expansion \cite{Riess,Perlmutter,Suzuki}, the phenomenon of dark energy and dark matter, which are essential in galactic dynamics, as well as the expansion of the size of the universe in accelerated mode, etc., are not adequately explained within the framework of GR without introducing exotic forms of matter and energy. Additionally, GR predicts singularities, such as those at the centres of black holes and the Big Bang, where the laws of physics break down. These singularities indicate a need for a more comprehensive theory that can be seamlessly integrated with quantum mechanics for a deeper understanding at the minute level. In this context, modified theories of gravity offer an intricate way to address these challenges by extending the geometrical and dynamical foundations of GR, potentially offering alternative explanations for these enigmatic phenomena and providing a more complete understanding of the universe. 

Soon after the advent of GR, Weyl \cite{Weyl} imposed the first modifications on it by introducing invariants of higher order to the action of Einstein-Hilbert to unify the electromagnetism and gravitation. Later, GR was modified by two types of geometrical extensions, {\it viz.},
\begin{itemize}
	\item extensions based on curvature, such as the $f(R),~f(T),~f(R,T)$, etc., theories of gravity.
	\item extensions based on torsion and non-metricity \cite{Einstein} such as the $f(Q)$ theory of gravity.
\end{itemize}
Such torsion and non-metricity based theories of gravity, which are equivalent to GR, and, are termed the Teleparallel Equivalent of General Relativity (TEGR) \cite{Hayashi,Sauer} and the Symmetric Teleparallel Equivalent of General Relativity (STEGR) \cite{Jimenez,Jimenez1}. In the context of flat space-time with torsion, the TEGR is formulated using the tetrads and spin connections as the primary variables. The constraints imposed by the nullification of the curvature and non-metricity tensor restrict the spin connection. This allows for the adoption of the Weitzenb\"ock connection, where all elements of the spin connections are eliminated, leaving only the tetrad components as the fundamental entities. This selection is interpreted as a gauge choice within the framework of TEGR. It is noteworthy that this choice does not influence the physical implications of the theory, as any other selection compatible with the teleparallelism constraint would yield the same action, barring a surface term. On the other hand, in the framework of STEGR, it is the non-metricity that characterises gravity, as opposed to curvature and torsion. Within the constraints of teleparallelism, a particular gauge, known as the coincidence gauge, can be selected in this theory. This gauge selection sets the metric tensor as the sole fundamental variable. An additional extension of the STEGR is the $f(Q)$ gravity theory, which bears a significant resemblance to the $f(R)$ gravity theory \cite{Jimenez,Heisenberg}. Within the general formalism of $f(Q)$ gravity, Hohmann et al. \cite{Hohmann} studied the potential polarisation and the velocity of propagation of gravitational waves in Minkowski space. Subsequently, Soudi et al. \cite{Soudi} showed that gravitational wave polarisation has an important impact on the strong-field nature of theories of gravity. $f(Q)$ theory has been utilised in a wide range of topics, including late-time acceleration \cite{Lazkoz}, black holes \cite{Ambrosio,Heisenberg1,Calza}, bouncing \cite{Bajardi}, evolution of the growth index in the context of matter fluctuations \cite{Khyllep}, and $f(Q)$ parametrisation to integrate the observational constraints \cite{Ayuso}. A detailed review of $f(Q)$ formalism has been described in Ref. \cite{Heisenberg2}. In addition, several aspects of $f(Q)$ gravity have been described in Refs.~\cite{Barros,Jimenez2,Anagnostopoulos,Flathmann,Heisenberg3}.  

In the astrophysical context, the study of compact objects within the framework of modified gravity is essential for advancing our understanding of fundamental physics. Compact objects such as black holes and neutron stars provide extreme environments where the effects of gravity are exceptionally strong, offering unique testing grounds for theories that extend beyond GR. These objects can reveal discrepancies between observed phenomena and theoretical predictions, shedding light on potential modifications to our current gravitational models. Recently, a large amount of data has been collected from pulsars and gravitational wave (GW) events and scrutinised through numerous research. This comprehensive analysis has led to precise predictions of several physical parameters pertaining to such celestial bodies. By utilising the Shapiro delay \cite{Cromartie} an accurate determination of the mass of millisecond pulsar PSR J0740+6620 resulted to be $2.14^{+0.10}_{-0.09}$. In addition to this, it is postulated that there could be a few compact celestial bodies that possess masses exceeding that of PSR J0740+6620 \cite{Cromartie}, particularly within the cluster of coupled systems. The occurrence of the event of gravitational wave, referred to as GW 190814, has indicated that the associated star boasts a mass ranging from 2.5 to 2.67 $M_{\odot}$, with a confidence level of 90\%. It is currently unclear whether it represents a massive compact star or a lightweight black hole. Accordingly, if the former is true, it becomes imperative to introduce a novel concept into the theoretical framework to extend the maximum mass range to account for such elevated mass values. This kind of formalism would be crucial for investigating the internal structure of dense matter beyond saturation density. 

The pressure anisotropy within a compact celestial body can be attributed to various factors. One potential cause is the existence of a superfluid of type 3A within the solid core of these objects \cite{Kippenhahn}. Due to the property of fermions, nucleons cannot occupy the same energy state simultaneously. Additionally, strong repulsive interactions between nucleons occur within very short ranges. These repulsive forces aid in counteracting gravitational collapse within a neutron star. However, this scenario changes when temperatures are extremely low, leading to the formation of Cooper pairs \cite{Broglia} of nucleons, which essentially behave as bosons. Consequently, at very low temperatures, nucleons exhibit collective behaviour on a large scale. These assembly of nucleons may condensate to form a state analogous to Helium-3, and, can flow without any viscosity. The high-pressure region, within a neutron star, effectively raises the critical temperature for the formation of Cooper pair, allowing for the occurrence of nuclear superfluidity even at temperatures of the order of a billion degrees. There could be three distinct types of superfluids present within the core of a neutron star \cite{Page}. Other potential sources of anisotropy inside compact objects could include the phase transitions \cite{Sokolov}, pion condensation \cite{Sawyer}, slow rotation and viscosity \cite{Herrera1}.

Considering electric charge in the study of compact stars is crucial because intense gravitational fields in such stars can separate charges, leading to a significant electric field. This electric charge can alter the star's structure, stability, and mass-radius relationship. Including charge in models may reveal new insights about high-energy phenomena, quark matter interactions, and the internal dynamics of stars, especially in extreme conditions found in NS. Rosseland \cite{Rosseland} was the first to propose that a star could be composed of a substantial number of electrons and positive ions. Given their extremely high kinetic energy, electrons have a higher likelihood of escaping from the star compared to positive ions. As a result, a star can possess a considerable amount of net positive charge. Such process persists until the internal electric field prevents further electron escape \cite{Eddington}. The Einstein Field Equations (EFE), when considering an electric field, are crucial for examining the viable properties of compact stars such as neutron stars, quark stars, strange stars, gravastars, black holes, and other entities in the realm of relativistic astrophysics. This is because the electric field significantly influences the overall properties of such compact objects. In his study, Ivanov \cite{Ivanov} showed that one of the most important prerequisites for obtaining a singularity-free perfect fluid distribution is the presence of non-vanishing net charge. While arguing on the results of Hoyle and Narlikar \cite{Hoyle} about the collapse of a large mass of gas into singularity, Bonnor \cite{Bonnor} showed that electric repulsion can counterbalance the gravitational collapse of a spherical body to achieve an equilibrium configuration. Stettner \cite{Stettner} showed that a spherical stellar structure in presence of net surface charge is found to be more stable than an uncharged one having a uniform density. Krasinski \cite{Krasinski} demonstrated that preventing the gravitational collapse of spherically distributed matter into a point singularity necessitates the presence of a net charge. Through a generalisation of the Oppenheimer–Volkoff equation, Bekenstein \cite{Bekenstein} conducted a qualitative assessment of the stability of spheres composed of charged fluid. Over the years, several researchers have addressed the impact of charge on compact stellar structures \cite{Saha,KBG2,Saha1,Saha2,Bhattacharjee,Pradhan,Singh,Pant,Maurya1}. 

Recently, the $f(Q)$ theory has garnered a significant amount of attention from researchers worldwide. In the astrophysical scenario, it has provided new insights into compact stellar structures. In the context of charged configurations, Kaur et al. \cite{Kaur} presented a charged $f(Q)$ solution to study an anisotropic fluid distribution using the Vaidya-Tikekar metric ansatz. Maurya et al. \cite{Maurya2} used an interesting approach of gravitational decoupling into the $f(Q)$ theory to study the constraining factors of mass and radii of the lighter component of GW190814 and other self-bound strange stars. Errehymy et al. \cite{Errehymy} studied a singularity-free charged strange star model in the framework of $f(Q)$ gravity. Furthermore, Lohakare et al. \cite{Lohakare} investigated the effects of gravitational decoupling on the maximum mass and stability of charged strange stars. In addition, several authors have included $f(Q)$ theory in their analysis of a stable stellar structure \cite{Bhar,Bhar1,Bhar2,Gul}.  

Compact objects are extremely dense, meaning they have incredibly strong gravitational fields. The introduction of charge into the system introduces an additional repulsive force due to electrostatic Coulomb interaction. The specific form of the charge distribution helps ensure that the object remains stable under the interplay of these forces balancing gravitational attraction with the repulsion caused by the charge. A uniform or non-uniform charge distribution can influence the overall stability, so the form must be chosen carefully to ensure physical plausibility. The charge distribution, when introduced, must be compatible with the EoS to avoid violations of the energy conditions and to ensure that the object behaves consistently with relativistic models. If the charge is distributed too densely in certain regions, it can lead to unphysical behaviours, such as superluminal sound speeds or collapse under excess electrostatic pressure.The inclusion of charge modifies the spacetime geometry around the compact object. The Einstein-Maxwell field equations govern how the gravitational field interacts with the electromagnetic field generated by the charge. A specific form of charge ensures that the resulting geometry remains consistent with observed astrophysical phenomena. For example, a too concentrated charge could lead to the formation of horizons or singularities, which might be inconsistent with the object's intended physical properties. Keeping all these points in mind and following the work of Gon\c calves and Lazzari \cite{Goncalves}, we have chosen a specific form of charge distribution, $q=q_{0}r^{3}$ in this present manuscript which ensures a compatible relativistic solution. Additionally, it must be noted that the chosen ansatz for the electric charge distribution represents the simplest model commonly applied in charge distribution studies.

In the present manuscript, the impact of charge has been investigated on the structure of a singularity-free compact star model within the framework of linear $f(Q)$ action. The advantage of $f(Q)$ gravity lies in its ability to derive the equations of GR without the necessity of employing the affine connection, thereby enhancing the inertial gravitational acceleration. This intriguing aspect coupled with the charged configuration reveals interesting stellar properties. Solving TOV equations, we have predicted the possible maximum mass and radius of compact stars in the framework of $f(Q)$ gravity in the presence of charge and we note some interesting results. 

The paper is structured as follows: Section \ref{sec2} demonstrates the fundamental ground work of $f(Q)$ theory in the presence of charge. Section \ref{sec3} addresses the formulation and solution of the Einstein-Maxwell equations in the framework of $f(Q)$ gravity. The analytical expressions of the thermodynamic quantities like energy density $(\rho)$ and pressures (radial $(p_{r}$) and tangential $(p_{t})$) are obtained in this section. The effects of $f(Q)$ parameters on the boundary conditions are described in Section \ref{sec4}. Section \ref{sec5} deals with the physical characteristic features of the model along with the determination of the maximum mass and radius, causality condition and energy conditions. The necessary stability criteria are addressed in Section \ref{sec6}. Finally, we conclude by summarising the major findings of the paper in Section \ref{sec7}. 
\section{Fundamentals of $f(Q)$ gravity in the presence of charge} \label{sec2} In this framework of symmetric teleparallel $f(Q)$ gravity, the gravitational action integral in the presence of an electromagnetic field is expressed as \cite{Jimenez,Zhao}: 
\begin{equation}
	\mathfrak{S}=\int \sqrt{-g}d^{4}x\Bigg[\frac{1}{2}f(Q)+\lambda^{kij}_{l}R^{l}_{kij}+\tau^{ij}_{k}T^{k}_{ij}+\mathfrak{L_{m}}+\mathfrak{L_{q}}\Bigg], \label{eq1}
\end{equation}
where, $g$ defines the determinant of the fundamental metric tensor $g_{ij}$, i.e., $g=|g_{ij}|$, $f(Q)$ is a function of $Q$ known as non-metricity, $\lambda^{kij}_{l}$ and $\tau^{ij}_{k}$ are two Lagrangian multipliers, $R^{l}_{kij}$ and $T^{k}_{ij}$ are the Riemann tensor and torsion tensor, respectively, and $\mathfrak{L_{m}}$ and $\mathfrak{L_{q}}$ describe the Lagrangian density of matter and electromagnetic charge, respectively. Using affine connections $(\Gamma^{k}_{ij})$, the non-metricity is expressed as: 
\begin{equation}
	Q_{kij}=\nabla_{k}g_{ij}=\delta_{k}g_{ij}-\Gamma^{l}_{ij}g_{ij}-\Gamma^{l}_{ik}g_{jl}, \label{eq2}
\end{equation} 
where, $\nabla_{k}$ is the covariant derivative. Again, the affine connection can be sub-divided into three components in the following way:
\begin{equation}
	\Gamma^{k}_{ij}=\epsilon^{k}_{ij}+K^{k}_{ij}+L^{k}_{ij}. \label{eq3} 
\end{equation}
In the above expression, 
\begin{itemize}
	\item $\epsilon^{k}_{ij}$ is the Levi-Civita connection which, with the help of the fundamental metric tensor $g_{ij}$, is defined as: 
	\begin{equation}
		\epsilon^{k}_{ij}=\frac{1}{2}g^{kl}\Big(\partial_{i}g_{lj}+\partial_{j}g_{il}-\partial_{l}g_{ij}\Big). \label{eq4}
	\end{equation}
	\item $K^{k}_{ij}$ describes the contorsion and is expressed as:
	\begin{equation}
		K^{k}_{ij}=\frac{1}{2}T^{k}_{ij}+T_{(_{i}~k~_{j})}. \label{eq5}
	\end{equation}
	In the STEGR, the contorsion ($K^{k}_{ij}$) physically represents the anti-symmetric part of the affine connection, which is written in the following form: $T^{k}_{ij}=2\Gamma^{k}_{[ij]}=\Gamma^{k}_{ij}-\Gamma^{k}_{ji}$. 
	\item In Eq.~(\ref{eq3}), the term $L^{k}_{ij}$ represents the deformation and is expressed as: 
	\begin{equation}
		 L^{k}_{ij}=\frac{1}{2}Q^{k}_{ij}+Q_{(_{i}~k~_{j})}. \label{eq6}
	\end{equation}
\end{itemize}
The superpotential associated with the non-metricity is described as,
\begin{equation}
	P^{kij}=-\frac{1}{4}Q^{kij}+\frac{1}{2}Q^{(ij)k}+\frac{1}{4}(Q^{k}-\tilde{Q}^{k})g^{ij}-\frac{1}{4}\delta^{k(_{i}Q_{j})}, \label{eq7}
\end{equation}
where, $Q^{k}$ and $\tilde{Q}^{k}$ define the independent traces of $Q_{kij}$ as:
\begin{equation}
	Q_{k}\equiv {Q_{k}}^{i}_{i}, ~~~~~~~~~~~ \tilde{Q}^{k}=Q_{i}^{ki}. \label{eq8}
\end{equation}
Ultimately, the non-metricity scalar is expressed by the following equation:
\begin{equation}
	Q=-g^{ij}\Big(L^{k}_{lj}L^{l}_{ik}-L^{l}_{il}L^{k}_{ij}\Big)=-Q_{kij}P^{kij}. \label{eq9}
\end{equation}
In absence of torsion as well as curvature, the affine connection can be effectively characterised by a tractable set of functions described as: 
\begin{equation}
	\Gamma^{k}_{ij}=\Bigg(\frac{\partial x^{k}}{\partial\chi^{l}}\Bigg)~ \partial_{i}\partial_{j}\chi^{l}, \label{eq10}
\end{equation}
 where, in the context of a general co-ordinate transformation, we consider arbitrary spacetime position functions denoted by $\chi^{l}$. Notably, we always have the flexibility of choosing the co-ordinate of the form $\chi^{l}=\chi^{l}(x^{i})$. Following the study of Jim\'enez et al. \cite{Jimenez}, the generality of affine connection, i.e., $\Gamma^{k}_{ij}=0$, is termed the coincident gauge property. Hence, the covariant derivatives in the coincident gauge reduce to ordinary derivatives in the standard GR, and consequently, the non-metricity formalism of Eq.~(\ref{eq3}) can be simplified as:
 \begin{equation}
 	Q_{kij}=\partial_{k}g_{ij}. \label{eq11}
 \end{equation}
Now, we obtain the tractable set of gravitational field equations by varying the Einstein-Hilbert action as written in Eq.~(\ref{eq1}) in terms of the metric tensor $g_{ij}$ of the form given below: 
\begin{equation}
	\frac{2}{\sqrt{-g}}\nabla_{k}\Big(\sqrt{-g}f_{Q}P^{k}_{ij}\Big)+\frac{1}{2}g_{ij}f+f_{Q}\Big(P_{ikl}Q^{kl}_{j}-2Q_{kli}P^{kl}_{j}\Big)=-\Big(T_{ij}+E_{ij}\Big), \label{eq12}
\end{equation}
where, $f_{Q}=\frac{\partial f}{\partial Q}$, $T_{ij}$ and $E_{ij}$ respectively represent the energy-momentum tensors of the matter and electromagnetic contribution. In a generalised way, we can describe $T_{ij}$ and $E_{ij}$ in the following forms:
\begin{equation}
	T_{ij}=-\frac{2}{\sqrt{-g}}\frac{\delta(\sqrt{-g}\mathfrak{L_{m}})}{\delta g^{ij}} ~~~~~~~and~~~~~~~~~E_{ij}=-\frac{2}{\sqrt{-g}}\frac{\delta(\sqrt{-g}\mathfrak{L_{q}})}{\delta g^{ij}}. \label{eq13}
\end{equation}
Moreover, the variation of action given in Eq.~(\ref{eq1}) with respect to the affine connection yields, 
\begin{equation}
	\nabla_{i}\nabla_{j}\Big(\sqrt{-g}f_{Q}P^{k}_{ij}-\frac{1}{2}\frac{\delta\mathfrak{L_{m}}}{\delta\Gamma^{k}_{ij}}\Big)=0, \label{eq14} 
\end{equation} 
where, the second term in Eq.~(\ref{eq14}) defines the tensor density of the hyper momentum. Imposing the relation $\nabla_{i}\nabla_{j}(H^{k}_{ij})=0$, Eq.~(\ref{eq14}) reduces to $\nabla_{i}\nabla_{j}(\sqrt{-g}f_{Q}P^{k}_{ij})=0$.
\section{Mathematical formalism: $f(Q)$ gravity and Einstein-Maxwell field equations} \label{sec3} In this paper, we investigate a novel generalised approach to charged compact star modeling in the framework of the $f(Q)$ theory of gravity. Within the realm of gravitational theories, the concept of static spherically symmetric spacetime serves as a foundational assumption, offering valuable insights into various aspects of astrophysical phenomena. Keeping this in mind, a static spherically and symmetric spacetime has been considered for the analysis of the present paper. Now, in the static spherically symmetric curvature co-ordinate space $(t,r,\theta,\phi)$, the line element is expressed as: 
\begin{equation}
	ds^2=-e^{2\nu(r)}dt^2+e^{2\lambda(r)}dr^2+r^2(d\theta^2+sin^2\theta d\phi^2). \label{eq15}
\end{equation}
Using Eq.~(\ref{eq15}) in Eq.~(\ref{eq9}), we obtain the expression of non-metricity scalar as:
\begin{equation}
	Q=-\frac{2e^{-2\lambda(r)}}{r}\Big(2\nu'(r)+\frac{1}{r}\Big), \label{eq16}
\end{equation}
where, the prime denotes the derivative with respect to the radial co-ordinate $(r)$ and, interestingly, $Q$ relies on the mathematical structure where the parallel transport, around any closed loop or in the Q-geometry, has no rotations or distortions. In other words, $Q$ is based upon null coefficients of affine connections. Now, the anisotropic perfect fluid distribution is characterised as: 
\begin{equation}
	T_{ij}=(\rho+p_{t})u_{i}u_{j}+p_{t}g_{ij}+(p_{r}-p_{t})v_{i}v_{j}, \label{eq17}
\end{equation}       
where, $u_{i}$ represents the four velocity related to the anisotropic perfect fluid and $v_{i}$ represent the space-like unit vector along the radial direction. These parameters obey the relations $u^{i}u_{i}=-1$ and $v^{i}v_{i}=1$. The terms $\rho,~p_{r},~p_{t}$ in Eq.~(\ref{eq17}) represent respectively the energy density and pressures along the radial and tangential directions associated with the fluid. The electromagnetic energy-momentum tensor takes the form:
\begin{equation}
	E_{ij}=\frac{1}{4\pi}\Big(F^{\eta}_{i}F_{j\eta}-\frac{1}{4}g_{ij}F_{\zeta\eta}F^{\zeta\eta}\Big). \label{eq18}
\end{equation} 
Here, $F_{\zeta\eta}$ represents the Faraday-Maxwell tensor of the form, $F_{\zeta\eta}=\partial_{\zeta}A_{\eta}-\partial_{\eta}A_{\zeta}$, where $A_{\zeta\eta}$ represents the electromagnetic four potential. The fundamental Maxwell equations governing electromagnetism are represented as:
\begin{equation}
	(\sqrt{-g}F^{ij})_{,j}=4\pi J^{i}\sqrt{-g}, ~~~~~~~~~~~F_{[ij,\delta]}=0. \label{eq19}
\end{equation} 
In the above expression, $J^{i}=\sigma u^{i}$, represents the electric four current, and $\sigma$ defines the electric charge density. Following, Maxwell's equations, the strength of the electric field is expressed as:
\begin{equation}
	E(r)=\frac{e^{\nu+\lambda}}{r^{2}}q(r), \label{eq20}
\end{equation}
where, $q(r)$ in Eq.~(\ref{eq20}) represents the total charge associated with the fluid sphere of radius $r$ and is written as:
\begin{equation}
	q(r)=4\pi\int^{r}_{0}\sigma r^{2}e^{\lambda}dr. \label{eq21}
\end{equation}
Conversely, the charge density can be written as: 
\begin{equation}
	\sigma=\frac{e^{-\lambda}\frac{dq(r)}{dr}}{4\pi r^{2}}. \label{eq22}
\end{equation}
To examine the impact of electric charge on the structure of the present model, it is essential to define the charge distribution within the star. We assume that the electric charge varies according to a specific function, dependent on the radial coordinate. In the present analysis, we have considered a particular form of charge as follows \cite{Goncalves}:
\begin{equation}
	q=q_{0}r^{3}, \label{eq23}
\end{equation}
where, $q_{0}$ is the intensity of charge having dimension of $Km^{-2}$ and $q_{0}=0$ describes a neutral uncharged case. 

Following the equation of motion given in Eq.~(\ref{eq12}) along with the anisotropic perfect fluid distribution of Eq.~(\ref{eq17}), we obtain a tractable set of the non-zero components of the Einstein-Maxwell field equations as follows:
\begin{eqnarray}
	\frac{f(Q)}{2}-f_{Q}\Big[Q+\frac{1}{r^{2}}+\frac{2e^{-2\lambda}}{r}(\nu'+\lambda')\Big]=8\pi\rho+\frac{q^{2}}{r^{4}}, \label{eq24} \\
	-\frac{f(Q)}{2}+f_{Q}\Big[Q+\frac{1}{r^{2}}\Big]=8\pi p_{r}-\frac{q^{2}}{r^{4}}, \label{eq25} \\
	-\frac{f(Q)}{2}+f_{Q}\Big[\frac{Q}{2}-e^{-2\lambda}\Big\{\nu''+2~\Big(\frac{\nu'}{2}+\frac{1}{2r}\Big)(\nu'-\lambda')\Big\}\Big]=8\pi p_{t}+\frac{q^{2}}{r^{4}}, \label{eq26} \\
	\frac{cot\theta}{2}Q'f_{QQ}=0, \label{eq27}
\end{eqnarray}
where, $(')$ denotes derivatives with respect to $r$. Now, Eq.~(\ref{eq27}) is utilised to obtain the form of linear $f(Q)$ action and is expressed as: 
\begin{equation}
	f(Q)=\alpha_{0}+\alpha_{1}Q, \label{eq28}
\end{equation}
where, $\alpha_{0}$ has the dimension of $Km^{-2}$ and $\alpha_{1}$ is dimensionless. Now, using Eqs.~(\ref{eq24}), (\ref{eq25}), (\ref{eq26}), (\ref{eq27}) and (\ref{eq28}), the field equations in the modified $f(Q)$ theory of gravity reduce to an exact set of formulations in the following form:
\begin{eqnarray}
	\frac{1}{2r^{2}}\Big[r^{2}\alpha_{0}-2\alpha_{1}e^{-2\lambda}(2r\lambda'-1)-2\alpha_{1}\Big]=8\pi\rho+\frac{q^{2}}{r^{4}}, \label{eq29}\\
	\frac{1}{2r^{2}}\Big[-r^{2}\alpha_{0}-2\alpha_{1}e^{-2\lambda}(2r\nu'+1)+2\alpha_{1}\Big]=8\pi p_{r}-\frac{q^{2}}{r^{4}}, \label{eq30} \\
	\frac{e^{-2\lambda}}{2r}\Big[-r\alpha_{0}e^{2\lambda}-2r\alpha_{1}\nu''-2\alpha_{1}(r\nu'+1)(\nu'-\lambda')\Big]=8\pi p_{t}+\frac{q^{2}}{r^{4}}. \label{eq31}
\end{eqnarray} 
Choice of metric potentials $\lambda(r)$ and $\nu(r)$ are very crucial as the components of gravitational field equations depend on them. Lake \cite{Lake} proposed that $\lambda(r)$ should be finite and constant at the stellar centre $(r=0)$ as well as must hold the following properties: $\lambda'(0)=0$ and $\lambda''(0)>0$. On the other hand, near the core, $\nu(r)$ must have the form $\nu=1+\mathcal O(r^{2})$, and it must increase towards the stellar boundary. Such behaviour of the metric potentials, $\lambda(r)$ and $\nu(r)$ guarantees a monotonically decreasing energy density and pressure profile. This characteristic is crucial for a stable and physically realistic stellar system. To fulfil the specified constraints in the present model, we opt for the Krori-Barua (KB) metric ansatz \cite{KB} in the form $\lambda=Ar^{2}$ and $\nu=Br^{2}+C$. Now, plugging the metric potentials along with Eq.~(\ref{eq23}) in the set of equations Eqs.~(\ref{eq29}), (\ref{eq30}) and (\ref{eq31}), we obtain the non-zero components of the field equations as:
\begin{equation}
	\rho=\frac{2\alpha_{1}e^{-2Ar^{2}}-2q_{0}^{2}r^{4}-2\alpha_{1}-r^{2}(8A\alpha_{1}e^{-2Ar^{2}}-\alpha_{0})}{16\pi r^{2}}, \label{eq32}
\end{equation} 
\begin{equation}
	p_{r}=\frac{-2\alpha_{1}e^{-2Ar^{2}}+2q_{0}^{2}r^{4}+2\alpha_{1}+r^{2}(8A\alpha_{1}e^{-2Ar^{2}}-\alpha_{0})}{16\pi r^{2}}, \label{eq33}
\end{equation}
\begin{equation}
	p_{t}=\frac{e^{-2Ar^{2}}(4A\alpha_{1}-\alpha_{0}e^{2Ar^{2}}-2q_{0}^{2}r^{2}e^{2Ar^{2}}-8B\alpha_{1}+8ABr^{2}\alpha_{1}-8B^{2}r^{2}\alpha_{1})}{16\pi}. \label{eq34}
\end{equation}
The pressure anisotropy parameter $(\Delta)$ is expressed as:
\begin{equation}
	\Delta=p_{t}-p_{r}. \label{eq35}
\end{equation}
Now, the total active gravitational mass in presence of net charge contained within a spherical volume of radius $R$ is, 
\begin{equation}
	m(r)=4\pi\int_{0}^{R}\Big(\rho+\frac{q^{2}}{8\pi r^{4}}\Big)~r^{2}~dr. \label{eq36}
\end{equation} 

\section{Boundary conditions}\label{sec4}
In the context of modified theories of gravity, the form of equations of motion as prescribed by GR are modified. Therefore, the associated boundary conditions applicable to any stellar system in GR require suitable modifications \cite{Rosa}. Now, in the 4-dimensional spacetime manifold $(\Omega)$, there exist two distinct regions, {\it viz.}, the interior and exterior space-times. These regions are demarcated by a 3-dimensional hypersurface $(\Sigma)$ defined by the induced metric $h_{ij}$ in the $X^{i}$ coordinate system, with indices excluding the direction orthogonal to $\Sigma$. We express the projection tensors and the normal vector from $\Omega$ and $\Sigma$ in the form, $e^{a}_{i}=\frac{\partial x^{a}}{\partial X^{i}}$ and $n_{a}=\epsilon\delta_{a}\ell$, where, the affine connection in the perpendicular direction to $\Sigma$ is denoted by $\ell$ and for the respective timelike, null and spacelike geodesics, $\epsilon$ takes the values $-1,~0$ or $1$. In light of the above composition, one considers $n^{a}e^{i}_{a}=0$. Now, the induced metric $h_{ij}$ and the extrinsic curvature tensor $K_{ij}$ at the hypersurface are written in the form:
\begin{equation}
	h_{ij}=e^{a}_{i}e^{b}_{j}g_{ab}, ~~~~~~~~~~~~~~~~~~~~~~K_{ij}=e^{a}_{i}e^{b}\nabla_{a}n_{b}. \label{eq37}
\end{equation}
In the framework of the distribution formalism introduced by Rosa et al. \cite{Rosa1}, one considers $[h_{ij}=0]$, which reflects the continuity of the induced metric at $\Sigma$. However, to determine the constants appearing in the metric potentials, we must ensure that the extrinsic curvature tensor is also continuous at the junction. In the absence of a thin shell, the continuity of the extrinsic curvature tensor is written as $[K_{ij}=0]$ \cite{Rosa1}. Now, to compute the extrinsic curvature, we begin by considering the vacuum charged exterior Reissner-Nordstr\"om \cite{Reissner,Nordstrom} solution expressed as: 
\begin{equation}
	ds^{2}=-\Bigg(1-\frac{2M}{r}+\frac{Q_{c}^{2}}{r^{2}}\Bigg)~dt^{2}+\Bigg(1-\frac{2M}{r}+\frac{Q_{c}^{2}}{r^{2}}\Bigg)^{-1}dr^{2}+r^{2}(d\theta^{2}+sin^{2}\theta d\phi^{2}), \label{eq38}
\end{equation}
where $M=$ the total mass and $Q_{c}=$ the total charge of the stellar configuration of radius $R$ respectively. Interestingly, in the parametric space of spherically symmetric space-time, $K_{ij}^{\pm}$ contains only two components {\it viz.}, $K_{00}$ and $K_{\theta\theta}=K_{\phi\phi}sin^{2}\theta$, where the $(+)$ and $(-)$ signs are for the exterior and interior space-time respectively. Using the KB ansatz and Eq.~(\ref{eq38}), we obtain the following components:
\begin{equation}
	K_{00}^{+}=\frac{-M-2q_{0}^{2}r^{5}}{r^{2}\sqrt{\frac{r}{r-2M+q_{0}^{2}r^{5}}}} ~~~~~~~~~~~~~~~~~~~~~ K_{00}^{-}=-2Bre^{-Ar^{2}+2Br^{2}+2C}, \label{eq39}
\end{equation}
\begin{equation}
	K_{\theta\theta}^{+}=\frac{r}{\sqrt{\frac{r^{2}}{Q_{c}^{2}-2Mr+r^{2}}}}~~~~~~~~~~~~~~~~~~~~~K_{\theta\theta}^{-}=re^{-Ar^{2}}. \label{eq40}
\end{equation}
Another necessary condition is that, at the surface of the star, radial pressure vanishes, i.e., 
\begin{equation}
	p_{r}(r=R)=0. \label{eq41}
\end{equation}
Now, matching the extrinsic curvature tenors written in Eqs.~(\ref{eq39}) and (\ref{eq40}) at the stellar boundary $(r=R)$ and using Eq.~(\ref{eq41}), we obtain:
\begin{equation}
	A=\frac{ln\Big[\frac{R^{2}}{R^{2}-2MR+Q_{c}^{2}}\Big]}{2R^{2}}, \label{eq42}
\end{equation}
\begin{equation}
	B=\frac{2q_{0}^{2}R^{6}-\alpha_{0}R^{4}-2\alpha_{1}Q_{c}^{2}+4\alpha_{1}MR}{8\alpha_{1}R^{2}(Q_{c}^{2}-2MR+R^{2})}, \label{eq43}
\end{equation}
\begin{equation}
	C=\frac{1}{2}\Big(\frac{4M\alpha_{1}-\alpha_{0}R^{3}-2q_{0}^{2}R^{5}(\alpha_{1}-1)}{8M\alpha_{1}-4R\alpha_{1}(q_{0}^{2}R^{4}+1)}\Big)+ln\Big[\frac{R}{R-2M+q_{0}^{2}R^{5}}\Big]+ln\Big[-\frac{4\alpha_{1}(M+2q_{0}^{2}R^{5})(R-2M+q_{0}^{2}R^{5})^{2}}{\alpha_{0}R^{5}+2q_{0}^{2}R^{7}(\alpha_{1}-1)-4MR^{2}\alpha_{1}}\Big]. \label{eq44}
\end{equation}
\section{Physical characteristics of the proposed model} \label{sec5}
It is very crucial that any stellar model should be physically viable to ensure that the physical parameters associated with the star, obtained using the model, must be physically realistic. A stellar model must comply with the regularity and reality conditions persistent in the context of relativistic stellar modeling to qualify as a compact star. In this section, we have considered the compact star PSR J0740+6620 of estimated mass $2.072~M_{\odot}$ and radius $12.39~Km$ \cite{Riley} to study the impact of charge on various stellar parameters and to test the physical viability of the present theoretical model within the framework of modified $f(Q)$ gravity. We have noted that the choice of charge intensity $(q_{0})$ in the present model, in the range of $0.0002-0.0009~Km^{2}$ yields a viable stellar model and notably, following the study of Maurya et al. \cite{Maurya}, we have considered $\alpha_{0}=10^{-46}~Km^{-2}$ along with $\alpha_{1}=-0.5$ as arbitrary choices.
\subsection{\bf Thermodynamic properties: energy density, radial and tangential pressures and pressure anisotropy} 
\begin{figure}[h!]
\centering
\includegraphics[width=0.5\textwidth]{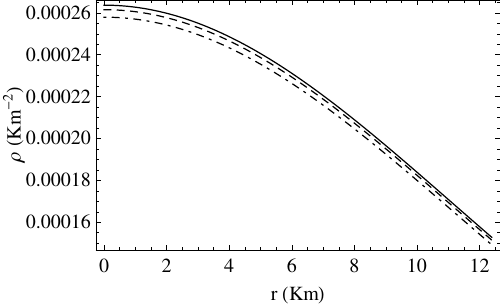}
\caption{Radial variation of energy density $(\rho)$ for different charge intensities $(q_{0})$. The solid, dashed and dot-dashed lines represent $q_{0}=0.0002,~0.0004$ and $0.0006~Km^{-2}$ respectively.}
\label{fig1}
\end{figure}

\begin{figure}[h!]
	\centering
	\includegraphics[width=0.5\textwidth]{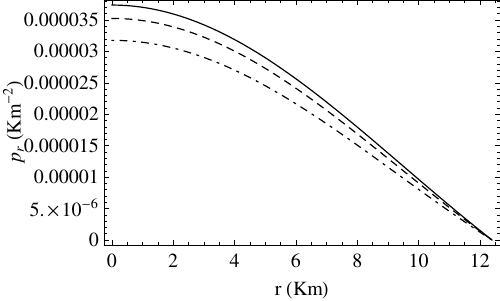}
	\caption{Radial variation of radial pressure $(p_{r})$ for different charge intensities $(q_{0})$. The solid, dashed and dot-dashed lines represent $q_{0}=0.0002,~0.0004$ and $0.0006~Km^{-2}$ respectively.}
	\label{fig2}
\end{figure}

\begin{figure}[h!]
	\centering
	\includegraphics[width=0.5\textwidth]{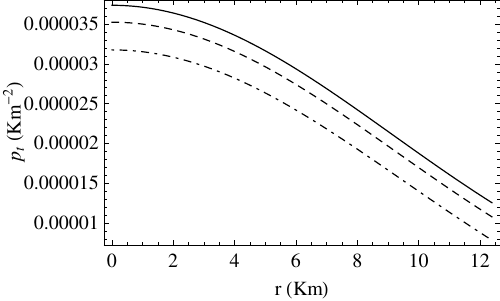}
	\caption{Radial variation of tangential pressure $(p_{t})$ for different charge intensities $(q_{0})$. Solid, dashed and dot-dashed lines correspond to $q_{0}=0.0002,~0.0004$ and $0.0006~Km^{-2}$ respectively.}
	\label{fig3}
\end{figure}

\begin{figure}[h!]
	\centering
	\includegraphics[width=0.5\textwidth]{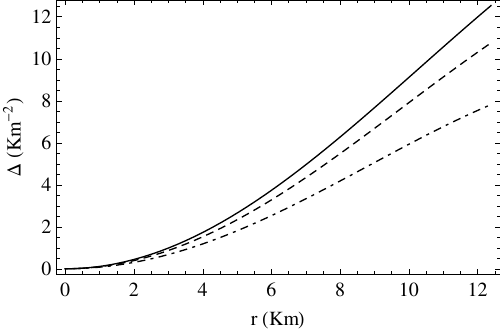}
	\caption{Radial variation of pressure anisotropy $(\Delta\times10^{-6})$ for different charge intensities $(q_{0})$. Solid, dashed and dot-dashed lines correspond to $q_{0}=0.0002,~0.0004$ and $0.0006~Km^{-2}$ respectively.}
	\label{fig4}
\end{figure}

To ensure a physically viable stellar model, the properties of energy density and pressures must have a nature of monotonically decreasing behaviour with r. Moreover, for a physically acceptable solution, the value of tangential pressure must be positive at all points interior of the compact star. The anisotropy in pressure, must be zero at the centre, i.e., $\Delta=0$, which reflects that $p_{r}=p_{t}$ at the stellar core. These claims are supported by Figs.~\ref{fig1}, \ref{fig2}, \ref{fig3} and \ref{fig4}. 
\begin{figure}[h!]
	\centering
	\includegraphics[width=0.5\textwidth]{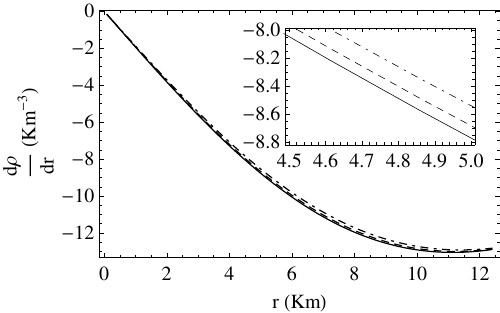}
	\caption{Radial variation of energy density gradient $(\frac{d\rho}{dr}\times10^{-6})$ for different charge intensities $(q_{0})$. The solid, dashed and dot-dashed lines correspond to $q_{0}=0.0002,~0.0004$ and $0.0006~Km^{-2}$ respectively.}
	\label{fig5}
\end{figure} 
\begin{figure}[h!]
	\centering
	\includegraphics[width=0.5\textwidth]{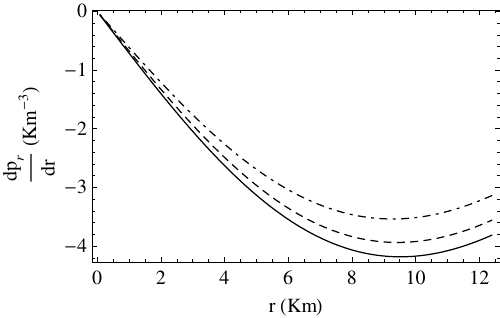}
	\caption{Radial variation of radial pressure gradient $(\frac{dp_{r}}{dr}\times10^{-6})$ for different charge intensities $(q_{0})$. The solid, dashed and dot-dashed lines represent $q_{0}=0.0002,~0.0004$ and $0.0006~Km^{-2}$ respectively.}
	\label{fig6}
\end{figure}
\begin{figure}[h!]
	\centering
	\includegraphics[width=0.5\textwidth]{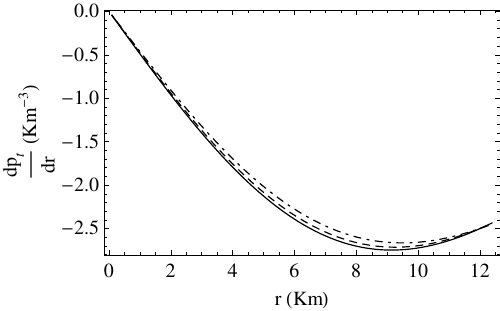}
	\caption{Radial variation of tangential pressure gradient $(\frac{dp_{t}}{dr}\times10^{-6})$ for different charge intensities $(q_{0})$. The solid, dashed and dot-dashed lines correspond to $q_{0}=0.0002,~0.0004$ and $0.0006~Km^{-2}$ respectively.}
	\label{fig7}
\end{figure}
From Figs.~\ref{fig1}, \ref{fig2} and \ref{fig3}, it is noted here that the physical parameters such as energy density $(\rho)$ and pressures $(p_{r}$ and $p_{t})$ attain a maximum value at the stellar core, and as the charge increases, these physical parameters decrease. This may be attributed to the fact that as charge increases within a compact star, the resulting electrostatic repulsion counteracts the inward gravitational pull, leading to a drop in both energy density and pressure profiles. This effect arises because the outward force generated by the accumulated charge reduces the compression needed to sustain high densities and pressures, especially in the inner regions. Consequently, the overall matter distribution adjusts, balancing the electric repulsion with gravitational forces, resulting in lower density and pressure profiles as shown in Figs.~\ref{fig1}, \ref{fig2} and \ref{fig3}. Moreover, the maximisation of the core density and pressure profiles is ensured by $\rho'(0)=0$, $p'_{r}(0)=0$ and $p'_{t}(0)=0$ and the nature of the gradients which is negative for all $0<r\leq R$. These behaviours of the gradients are shown in the Figs.~\ref{fig5}, \ref{fig6} and \ref{fig7}. Additionally, from Figs.~\ref{fig5}, \ref{fig6} and \ref{fig7}, we note that as the charge intensity $(q_{0})$ increases the profiles of the gradients increase. Moreover, Fig.~(\ref{fig4}) shows that $\Delta>0$ throughout the interior of the star and with increasing charge intensity $(q_{0})$, $\Delta$ decreases. Based on this behaviour, we may note that with increasing $q_{0}$, the electric field introduces an additional outward force which may help to equalise the pressure distribution by partially compensating for the intense inward gravitational pull, resulting in decreased pressure anisotropy. Furthermore, the positiveness of $\Delta$ denotes a repulsive anisotropic nature pointing towards a more massive structure. In light of the above physical analysis, we can say that the thermodynamic properties are well satisfied in this model, which points towards a physically acceptable stellar model in stable equilibrium.
\newpage
\subsection{\bf Maximum mass and radius from TOV equation} 
\begin{figure}[h]
	\centering
	\includegraphics[width=0.5\textwidth]{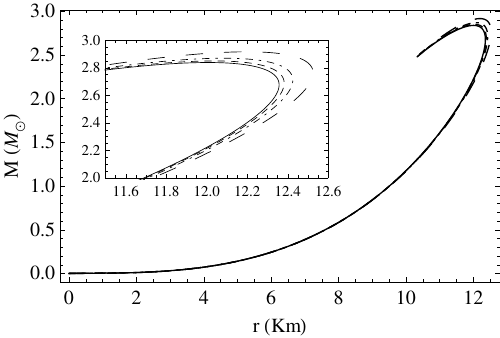}
	\caption{ Mass-radius plot for different charge intensities $(q_{0})$. The solid, dashed, dot-dashed and large-dashed lines represent $q_{0}=0.0002,~0.0004,~0.0006$ and $0.0009~Km^{-2}$ respectively.}
	\label{fig7a}
\end{figure}

In this section, we numerically solve the TOV equations \cite{Tolman,Oppenheimer} to obtain the maximum mass of a star as well as its radius for a charged compact star. It should be noted that, the solution of the TOV equation is entirely based upon the choice of a particular EoS. Hence, to resort to a suitable EoS, we adopt the method of maximising the radial sound velocity $(v_{r}^{2}=\frac{dp_{r}}{d\rho}=1)$, to the extreme end of causality, at the centre and the method of curve fitting from the data set of $\rho$ and $p_{r}$. We tabulate the EoS for increasing charge intensity $(q_{0})$ in Table~\ref{tab1}.   
\begin{table}[h]
\centering
\caption{Best fit EoS with increasing charge intensity}
\label{tab1}
\begin{tabular}{cc}
\hline
Charge intensity $(q_{0})$ $(Km^{-2})$ & EoS \\
\hline
0.0002 & $p_{r}=0.910646~\rho-0.000354486$\\
0.0004 & $p_{r}=0.910245~\rho-0.000354205$\\
0.0006 & $p_{r}=0.909595~\rho-0.000353745$\\
0.0009 & $p_{r}=0.908106~\rho-0.000352706$\\
\hline
\end{tabular}
\end{table}
Using Table~\ref{tab1}, we have numerically solved TOV equations and the results are plotted in Fig.~\ref{fig7a}. Maximum mass and corresponding radius are evaluated from Fig.~\ref{fig7a} and are tabulated in Table~\ref{tab2} for different charge intensities $(q_{0})$.  
\begin{table}[h]
	\centering
	\caption{Maximum mass and radius from TOV equation}
	\label{tab2}
	\begin{tabular}{ccc}
		\hline
		Charge intensity $(q_{0})$ $(Km^{-2})$ & Maximum mass $(M_{\odot})$ & Radius (Km)\\
		\hline
		0.0002 & 2.84 & 12.00\\
		0.0004 & 2.85 & 12.03\\
		0.0006 & 2.87 & 12.08\\
		0.0009 & 2.92 & 12.20\\
		\hline
	\end{tabular}
\end{table}
It is noted that maximum mass and radius both increase with charge intensity $(q_{0})$ in this model, which denotes the transition of the EoS from a softer to a stiffer nature.  We have also predicted the radii of some pulsars and GW events observed recently and evaluated their characteristic of physical parameters, {\it viz.}, central density $(\rho_{c})$, surface density $(\rho_{s})$ and central pressure $(p_{c})$, and they are tabulated in Table~\ref{tab3}.  
\begin{table}[h]
	\centering
	\caption{Tabulation of radius prediction and physical parameters}
	\label{tab3}
	\begin{tabular}{cccccccc}
		\hline
		Compact Object & Measured & Measured & Charge & Predicted & Central & Surface & Central\\
		& mass & radius & intensity & radius & density & density & pressure \\
		\vspace{0.25cm}
		& $(M_{\odot})$ & (Km) & $(q_{0})$ $(Km^{-2})$ & (Km) & $(\rho_{c})~(gm/cm^{3})$ & $(\rho_{s})~(gm/cm^{3})$ & $(p_{c})~(dyn/cm^{2})$ \\
		\hline
		\vspace{0.25cm}
		GW 190814 \cite{Abbott1} & $2.59^{+0.08}_{-0.09}$ & -- & 0.0006 & 11.82 & $0.41\times10^{15}$ & $0.23\times10^{15}$ & $0.52\times10^{35}$ \\
		\vspace{0.25cm}
		PSR J0952-0607 \cite{Carvalho1} & $2.35^{+0.17}_{-0.17}$ & -- & 0.0002 & 12.15 & $0.46\times10^{15}$ & $0.23\times10^{15}$ & $0.78\times10^{35}$ \\
		\vspace{0.25cm}
		PSR J0740+6620 \cite{Riley} & $2.072^{+0.067}_{-0.066}$ & $12.39^{+1.30}_{-0.98}$ & 0.0006 & 11.82 & $0.41\times10^{15}$ & $0.23\times10^{15}$ & $0.52\times10^{35}$ \\
		\vspace{0.25cm}
		4U 1820-30 \cite{Guver} & $1.58^{+0.06}_{-0.06}$ & $9.1^{+0.4}_{-0.4}$ & 0.0002 & 10.94 & $0.37\times10^{15}$ & $0.24\times10^{15}$ & $0.37\times10^{35}$ \\
		\vspace{0.25cm}
		EXO 1745-248 \cite{Ozel} & 1.4 & 11 & 0.0006 & 10.59 & $0.35\times10^{15}$ & $0.23\times10^{15}$ & $0.27\times10^{35}$ \\
		HER X-1 \cite{Abubekerov} & $0.85^{+0.15}_{-0.15}$ & $8.1^{+0.41}_{-0.41}$ & 0.0002 & 9.05 & $0.32\times10^{15}$ & $0.24\times10^{15}$ & $0.17\times10^{35}$ \\
		\hline
	\end{tabular}
\end{table}
\newpage
\subsection{\bf Causality condition}
\begin{figure}[h]
	\centering
	\includegraphics[width=0.5\textwidth]{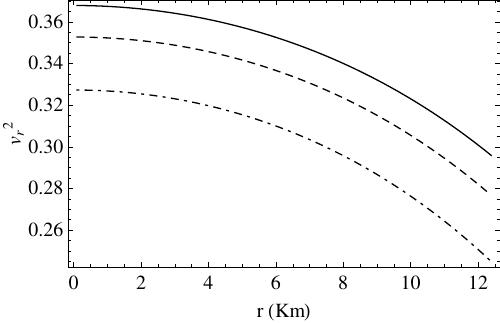}
	\caption{Radial variation of radial sound velocity $(v_{r}^{2})$ for different charge intensities $(q_{0})$. The solid, dashed and dot-dashed lines represent $q_{0}=0.0002,~0.0004$ and $0.0006~Km^{-2}$ respectively.}
	\label{fig8}
\end{figure}

\begin{figure}[h!]
	\centering
	\includegraphics[width=0.5\textwidth]{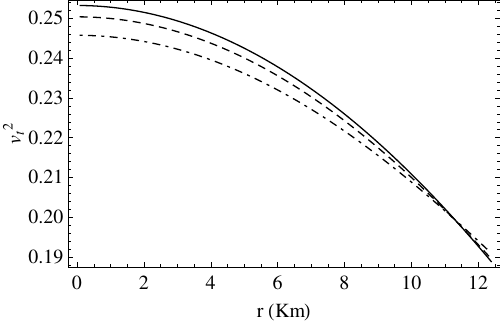}
	\caption{Radial variation of tangential sound velocity $(v_{t}^{2})$ for different charge intensities $(q_{0})$. The solid, dashed and dot-dashed lines represent $q_{0}=0.0002,~0.0004$ and $0.0006~Km^{-2}$ respectively.}
	\label{fig9}
\end{figure}

In the pursuit of a realistic model for an anisotropic compact star, a valuable approach for characterising the dense interior matter lies in the investigation of sound wave speeds. The radial and tangential sound speeds are expressed as $v_{r}^{2}=\frac{dp_{r}}{d\rho}$ and $\frac{dp_{t}}{d\rho}$, respectively, where $\rho, p_{r}$ and $p_{t}$ are defined earlier. In this formulation, we have used $\hbar=c=1$, and the causality criterion imposes an absolute upper bound of the sound velocities as $v_{r}^{2}\leq1$ and $v_{t}^{2}\leq1$. However, the thermodynamic equilibrium ensures that $v_{r}^{2}>0$ and $v_{t}^{2}>0$. Hence, the combined effects of both conditions, i.e., $0<v_{r}^{2}\leq1$ and $0<v_{t}^{2}\leq1$, must be simultaneously satisfied within the stellar configuration. To circumvent the mathematical intricacies, we have opted to represent the radial variation of sound velocities graphically in Figs.~\ref{fig8} and \ref{fig9}.
It is evident here that the causality conditions are well maintained in the stellar interior within the parameter space used here.
\newpage
\subsection{\bf Energy conditions}
\begin{figure}[h!]
	\centering
	\includegraphics[width=0.5\textwidth]{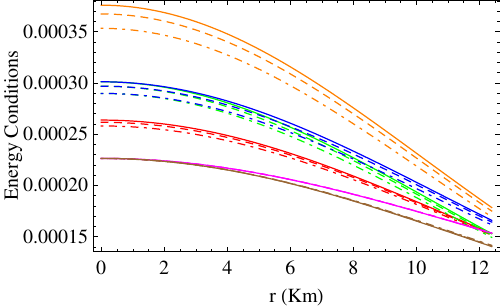}
	\caption{Radial variation of energy conditions for different charge intensities $(q_{0})$. Here, Red, Green, Blue, Orange, Magenta and Brown lines represent $(\rho+\frac{q^{2}}{8\pi r^{4}})$, $(\rho+p_{r})$, $(\rho+p_{t}+\frac{q^{2}}{4\pi r^{4}})$, $(\rho+p_{r}+2p_{t}+\frac{q^{2}}{4\pi r^{4}})$, $(\rho-p_{r}+\frac{q^{2}}{4\pi r^{4}})$ and $(\rho-p_{t})$ respectively. The solid, dashed and dot-dashed lines represent $q_{0}=0.0002,~0.0004$ and $0.0006~Km^{-2}$ respectively.}
	\label{fig10}
\end{figure}

In the field of gravitational theory, constraints known as energy conditions are applied to matter distributions. These constraints are essential for deriving a physically feasible energy-momentum tensor. Essentially, these conditions provide a method to explore the characteristics of distribution of matter without requiring detailed information about the composition of internal matter. Therefore, it is feasible to discern the physical characteristics of extreme phenomena, such as the presence of a geometrical singularity or gravitational collapse of merger, without explicit knowledge of pressure or energy density. Essentially, the examination of energy conditions is an algebraic issue \cite{Kolassis}, more precisely, it is an eigen value problem associated with the energy-momentum tensor. In a 4D spacetime, the investigation of energy conditions results in the roots of a quartic polynomial, a process that becomes complex due to the existence of analytical solutions for eigenvalues. Although finding a general solution can be challenging, a physically plausible fluid distribution should simultaneously adhere to the dominant energy, strong, weak and null conditions commonly known as the energy conditions \cite{Kolassis,Hawking,Wald} within the confines of the stellar configuration. In this study, we have examined the energy conditions \cite{Brassel,Brassel1} for the current stellar configuration in the form given below:
\begin{eqnarray}
	NEC:~\rho+\frac{q^{2}}{8\pi r^{4}}\geq0, \nonumber\\
	WEC:~\rho+p_{r}\geq0,~\rho+p_{t}+\frac{q^{2}}{4\pi r^{4}}\geq0, \nonumber\\
	SEC:~\rho+p_{r}+2p_{t}+\frac{q^{2}}{4\pi r^{4}}\geq0, \nonumber\\
	DEC:~\rho-p_{r}+\frac{q^{2}}{4\pi r^{4}}\geq0, \rho-p_{t}\geq0.\nonumber 
\end{eqnarray}
From Fig.~\ref{fig10}, we note that all the necessary energy conditions are satisfactorily obeyed throughout the stellar interior.
\section{Stability analysis}\label{sec6}
TO examine the stability of stellar structure in this model, we have applied the following methodologies:
\begin{itemize}
	\item TOV equation in a generalised form in presence of charge
	\item Herrera cracking condition
	\item Adiabatic index
	\item Zel'dovich condition
\end{itemize}
\subsection{\bf TOV equation in a generalised form in presence of charge} 
\begin{figure}[h]
	\centering
	\includegraphics[width=0.5\textwidth]{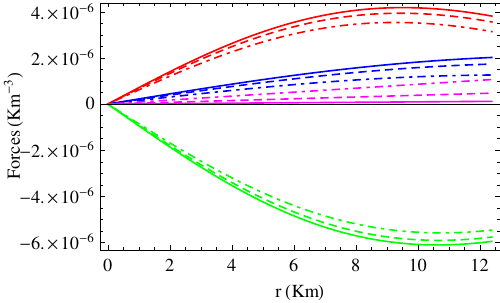}
	\caption{Radial variation of different forces for different charge intensities $(q_{0})$. Here, Green, Red, Blue and Magenta lines represent $F_{g}$, $F_{h}$, $F_{a}$ and $F_{q}$ respectively. The solid, dashed and dot-dashed lines represent $q_{0}=0.0002,~0.0004$ and $0.0006~Km^{-2}$ respectively.}
	\label{fig11}
\end{figure}

A model is physically realistic or not can be verified using the stability analysis subjected to various forces associated with the stellar configuration. For a charged compact star in presence of pressure anisotropy, this stability analysis focuses on four different components of forces: (i) gravitational force $(F_{g})$, (ii) hydrostatic force $(F_{h})$, (iii) anisotropic force $(F_{a})$, and (iv) electromagnetic force $(F_{q})$. To ensure that the model remains in stable equilibrium, the combined effects of these forces must be balanced. In this study, to analyse the stability, we utilise the generalised form of the TOV equation \cite{Tolman, Oppenheimer}, as expressed below:
\begin{equation}
	-\frac{M_{G}(\rho+p_{r})}{r^{2}}e^{\lambda-\nu}-\frac{dp_{r}}{dr}+\frac{2\Delta}{r}+\frac{q}{4\pi r^{4}}\frac{dq}{dr}=0. \label{eq45}
\end{equation}
The active gravitational mass, denoted by $M_{G}$ as given in Eq.~(\ref{eq45}) which can be derived using the mass formula as proposed by Tolman-Whittaker \cite{Gron} in the following form:
\begin{equation}
	M_{G}(r)=r^{2}\nu'e^{\nu-\lambda}. \label{eq46}
\end{equation}
Substituting Eq.~(\ref{eq46}) into Eq.~(\ref{eq45}), we obtain:
\begin{equation}
	-\nu'(\rho+p_{r})-\frac{dp_{r}}{dr}+\frac{2\Delta}{r}+\frac{q}{4\pi r^{4}}\frac{dq}{dr}=0. \label{eq47}
\end{equation}
Here, 
\begin{eqnarray}
	F_{g}=-\nu'(\rho+p_{r}), \label{eq48} \\
	F_{h}=-\frac{dp_{r}}{dr}, \label{eq49} \\
	F_{a}=\frac{2\Delta}{r}, \label{eq50}\\
	F_{q}=\frac{q}{4\pi r^{4}}\frac{dq}{dr}. \label{eq51}
\end{eqnarray}
By utilising Eqs.~(\ref{eq32}), (\ref{eq33}) and (\ref{eq34}) in Eqs.~(\ref{eq48}), (\ref{eq49}), (\ref{eq50}) and (\ref{eq51}), we obtain the equilibrium condition prescribed in the generalised TOV equations. To avoid the mathematical complexity, we have shown the condition of stable equilibrium through a graphical representation. Fig.~\ref{fig11} shows that our model is stable under the influence of different forces. 
\subsection{\bf Herrera cracking condition} 
\begin{figure}[h]
	\centering
	\includegraphics[width=0.5\textwidth]{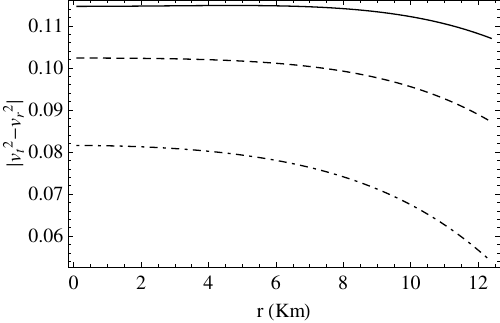}
	\caption{Radial variation of $|v_{r}^{2}-v_{t}^{2}|$ for different charge intensities $(q_{0})$. The solid, dashed and dot-dashed lines indicate that $q_{0}=0.0002,~0.0004$ and $0.0006~Km^{-2}$ respectively.}
	\label{fig12}
\end{figure}

According to Herrera \cite{Herrera}, the term "Cracking" within a fluid sphere of a self-gravitating compact object represents the total radial forces originated, each with different signs, when perturbations are introduced into the system. If the EoS is taken into account, cracking can arise from two scenarios, namely, i) a local anisotropic fluid, or ii) a slowly contracting and radiating perfect fluid. Utilising this cracking condition, Abreu et al. \cite{Abreu} explored the impact of local anisotropy on the distribution of local anisotropic fluid. They demonstrated that when perturbations exist in the system, potential stability can be established through the difference between the squares of radial $(v_{r}^{2})$ and tangential $(v_{t}^{2})$ sound velocities expressed as:
\begin{equation}
	0\leq|v_{t}^{2}-v_{r}^{2}|\leq1. \label{eq52}
\end{equation}
A stellar model that adheres to this relation is referred to as a stable structure. From Fig.~(\ref{fig12}), we note that the Abreu inequality is well maintained throughout the interior of the stellar configuration within the parameter space used here. 
\newpage
\subsection{\bf Adiabatic index}
\begin{figure}[h]
	\centering
	\includegraphics[width=0.5\textwidth]{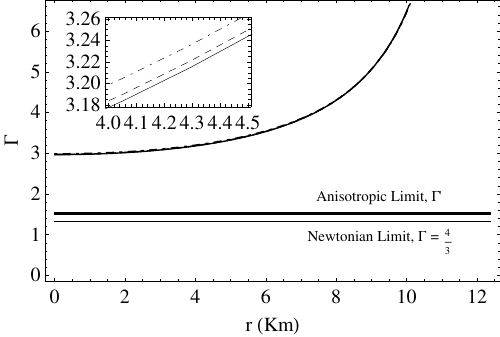}
	\caption{Adiabatic index $(\Gamma)$ vs r plot for different charge intensities $(q_{0})$. The solid, dashed and dot-dashed lines indicate $q_{0}=0.0002,~0.0004$ and $0.0006~Km^{-2}$ respectively.}
	\label{fig13}
\end{figure}

The adiabatic index $(\Gamma)$ signifies the rigidity of the EoS at a particular density. In the context of a relativistic anisotropic stellar structure, the adiabatic index is represented as:
\begin{equation}
	\Gamma=\frac{\rho+p_{r}}{p_{r}}\frac{dp_{r}}{dr}=\frac{\rho+p_{r}}{p_{r}}v_{r}^{2}. \label{eq53}
\end{equation} 
According to Heintzmann and Hillebrandt \cite{Heintzmann}, a Newtonian matter distribution demonstrates, $\Gamma>\frac{4}{3}$. Later, Chan et al. \cite{Chan} extended the upper bound of the adiabatic index considering the existence of pressure anisotropy, and it is articulated as follows:
\begin{equation}
	\Gamma>\Gamma', \label{eq54}
\end{equation}
where, 
\begin{equation}
	\Gamma'=\frac{4}{3}-\Bigg[\frac{4}{3}\frac{(p_{r}-p_{t})}{|p'_{r}|r}\Bigg]_{max}. \label{eq55}
\end{equation}
Now, to determine the anisotropic limit, we have evaluated $\Gamma'$ for different charge intensities $(q_{0})$ and the composite results are plotted in Fig.~\ref{fig13}. 
In our model, it is evident from Fig.~\ref{fig13} that the value of $\Gamma$ consistently exceeds the threshold limit $\Gamma'$ and hence the present model is stable under the consideration of adiabatic index variation.
\subsection{\bf Zel'dovich condition}
Based on the Zel’dovich condition \cite{Shapiro,Zeldovich,Zeldovich1,Zeldovich2}, a stellar structure is considered stable if the ratio of pressure to density is less than one throughout the entire structure. If we denote this ratio as $\Omega$, then considering the radial $(p_{r})$ and tangential $(p_{t})$ pressures, we obtain $\Omega_{r}=\frac{p_{r}}{\rho}$, and $\Omega_{t}=\frac{p_{t}}{\rho}$. The criterion for stability is that $\Omega_{r}$ and $\Omega_{t}$ must be less than 1.
\begin{figure}[h]
	\centering
	\includegraphics[width=0.5\textwidth]{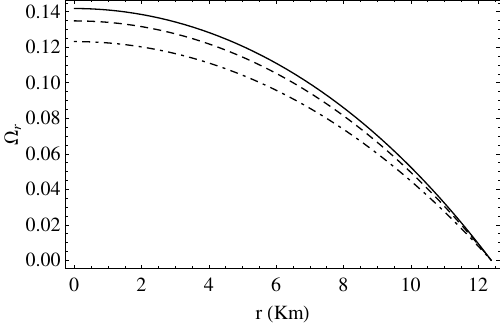}
	\caption{Radial variation of $(\Omega_{r})$ for different charge intensities $(q_{0})$. The solid, dashed and dot-dashed lines represent $q_{0}=0.0002,~0.0004$ and $0.0006~Km^{-2}$ respectively.}
	\label{fig14}
\end{figure}
\begin{figure}[h]
	\centering
	\includegraphics[width=0.5\textwidth]{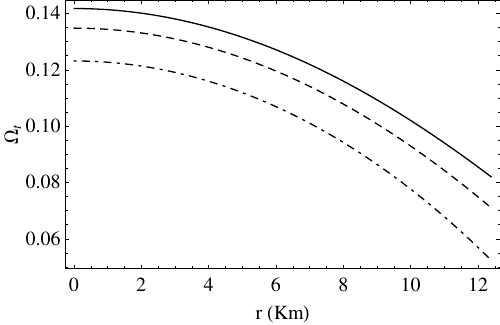}
	\caption{Radial variation of adiabatic index $(\Omega_{t})$ for different charge intensities $(q_{0})$. The solid, dashed and dot-dashed lines represent $q_{0}=0.0002,~0.0004$ and $0.0006~Km^{-2}$ respectively.}
	\label{fig15}
\end{figure}
It is evident from Figs.~\ref{fig14} and \ref{fig15} that the Zel'dovich criterion is well maintained within the stellar interior. Hence, the present model can be termed a stable configuration. 
\section{Discussion}\label{sec7}
Finally, we summarise the major findings of the present paper and our efforts to mimic a charged compact stellar configuration in the context of symmetric teleparallel $f(Q)$ gravity. Taking into account a spherically symmetric space-time, we combine the KB metric ansatz \cite{KB}, with a linear form of the $f(Q)$ action, which allows us to formulate a singularity-free solution of the EFE within the framework of $f(Q)$ theory. This solution incorporates a specific form of charge, denoted as $q(r)=q_{0}r^{3}$, where $q_{0}=0$ illustrates an uncharged stellar configuration.

Within this theoretical construct, we have selected PSR J0740+6620 \cite{Riley} as a potential candidate to study its properties in presence of charge. We analyse and represent our results, both numerically and graphically, by varying the charge intensity parameter $(q_{0})$. It has been shown in Refs. \cite{Rosa,Rosa1} that the modified theories of gravity lead to alterations in the boundary conditions of GR. This paper addresses these changes by assessing the continuity of both the induced metric as well as the extrinsic curvature tensors at the boundary of the stellar structure. We have used, (i) the matching condition of interior solutions with that of Reissner-Nordstr\"om \cite{Reissner,Nordstrom} exterior solution to compute the extrinsic curvature tensors, as obtained in Eqs.~(\ref{eq39}) and (\ref{eq40}), and (ii) the condition of nullification of radial pressure at the stellar surface, to compute the necessary constant present in the model. We have noted the following key characteristic features:
\begin{itemize} 
	\item Radial variation of the thermodynamic properties of a stellar structure, such as the energy density $(\rho)$, radial $(p_{r})$ and tangential $(p_{t})$ pressure profiles are represented in Figs.~\ref{fig1}, \ref{fig2}, \ref{fig3}. As the charge intensity increases, the energy density and pressure profiles decrease. However, the monotonically decreasing nature of the characteristic profiles is well preserved in this model. Additionally, we may note that the increase in charge within a compact star generates electrostatic repulsion, which acts against the inward gravitational force, leading to a decrease in both energy density and pressure profiles. This effect occurs because the outward force from the accumulated charge reduces the compressive requirement to maintain high densities and pressures, particularly in the star's core. As a result, the matter distribution shifts to balance the electric repulsion and gravitational forces, yielding lower density and pressure profiles, as illustrated in Figs.~\ref{fig1}, \ref{fig2}, and \ref{fig3}. Furthermore, the radial variation of anisotropy is illustrated in Fig.~\ref{fig4}. Fig.~\ref{fig4} indicates that as the charge intensity $(q_{0})$ increases, the electric field generates an additional outward force that aids in balancing the pressure distribution. This outward force partially offsets the strong inward gravitational pull, thereby reducing pressure anisotropy. Moreover, from Fig.~\ref{fig4}, we note that, $\Delta>0$, which indicates a repulsive anisotropic behaviour. This nature further indicates a more massive stellar structure. We also note that at the centre $(r\rightarrow0)$, $\Delta\rightarrow0$, i.e., $p_{r}=p_{t}$, which is another viable condition in favour of stable compact star modeling. The radial variations of pressure and density gradients are negative, as shown in Figs.~\ref{fig5}, \ref{fig6} and \ref{fig7}, which aids in the maximisation of the core density. Hence, it may be concluded that the physical parameters are consistent throughout the stellar structure.
	\item Obtaining the maximum mass through TOV equations depends on the particular choice of the EoS. In the present scenario, we have obtained the suitable EoS by maximising the radial sound velocity at the stellar centre, i.e., $\frac{dp_{r}}{d\rho}|_{r=0}=1$. Moreover, we have obtained the best fit EoS in this model, which are tabulated in Table~\ref{tab1} for different values of charge intensity $(q_{0})$. Using the EoS as tabulated in Table~\ref{tab1}, we have numerically evaluated the maximum mass by solving the TOV equations, and Fig.~\ref{fig7a} depicts the mass-radius plot. The maximum masses and radii for different charge intensities are listed in Table~\ref{tab2}. Solution of TOV equation in this present model predicts that the maximum mass attainable in this context is $2.92~M_{\odot}$ for charge intensity $q_{0}=0.0009~Km^{-2}$. Moreover, in this model the radii of different pulsars and lighter component of GW190814 event have been predicted and are tabulated in Table~\ref{tab3}. It is noted that our model complies with the observational results as well as predictions from other theoretical models \cite{KBG,KBG1,Rohit,Carvalho,Bhattacharjee1,Maurya3}. As for example, the predicted radius of lighter component of GW190814 event from our model is 11.82 Km which is very close to the value of $11.76^{+0.14}_{-0.19}$ Km predicted by Maurya et al. \cite{Maurya3} using the Einstein-Gauss-Bonnet theory of gravity. It is noted that the model is valid up to the value of charge intensity $q_{0}\leq0.0009~Km^{-2}$. Beyond this value, the model does not permit physically viable results.       
	\item Figs.~\ref{fig8} and \ref{fig9} show the variations of radial $(v_{r}^{2})$ and tangential $(v_{t}^{2})$ sound velocities with the radial distance.Notably, with increasing charge intensity, both $v_{r}^{2}$ and $v_{t}^{2}$ decrease within the stellar boundary. However, it is evident that both the velocities are well-regulated, as they do not surpass the speed of light and comply with the necessary causality condition within the constrained value of parameters.
	\item Considering the presence of charge, we have studied the necessary and sufficient energy conditions for a well-behaved compact star model within the framework of $f(Q)$ gravity, which is demonstrated in Fig.~\ref{fig10}. We note that, all the energy conditions are fulfilled within the stellar structure for the parameter space used here. 
	\item We have assessed the stability of the present model, within the charged $f(Q)$ formalism, using the generalised TOV approach, the cracking condition proposed by Herrera, the radial variation of the adiabatic index and the Zel'dovich criterion. Fig.~\ref{fig11} shows that the proposed model is in hydrostatic equilibrium under the impact of different forces and with the variation of charge intensity $(q_{0})$. To simulate a realistic stellar structure, perturbations must be considered within the system. With the presence of such perturbations, the potential stability of the charged model is obtained through the study of cracking condition, as proposed by Herrera. Fig.~\ref{fig12} shows that the cracking condition is well obeyed in this model. The modifications introduced in the adiabatic index due to the presence of anisotropy are written in Eq.~(\ref{eq55}) and the investigation of stability on the basis of the adiabatic index is shown in Fig.~\ref{fig13}. Figs.~\ref{fig14} and \ref{fig15} show that the Zel'dovich criterion is well satisfied throughout the charged stellar structure.  
\end{itemize}
To sum up, it is fascinating to observe that the $f(Q)$ gravity facilitates the creation of physically plausible models that align with fundamental principles in the context of static, spherically symmetric space-times. Therefore, considering all the key aspects of the present model, we can evidently state that we have presented here a stable and physically acceptable singularity-free charged compact star model within the framework of the $f(Q)$ theory of gravity.


	

\section{Acknowledgments}
DB is thankful to the Department of Science and Technology (DST), Govt. of India, for providing the fellowship vide no:  DST/INSPIRE Fellowship/2021/IF210761. PKC gratefully acknowledges support from IUCAA, Pune, India under Visiting Associateship programme.

\appendix

 \bibliographystyle{elsarticle-num} 





\end{document}